\documentclass{aastex61}

\shorttitle{The Cepheid metallicity effect in VIJHK bands}
\shortauthors{Wielg{\'o}rski et al.}

\begin{document}

\title{A precision determination of the effect of metallicity on Cepheid
absolute magnitudes in VIJHK bands from Magellanic Cloud Cepheids}

\correspondingauthor{Piotr Wielg{\'o}rski}
\email{pwielgor@camk.edu.pl}

\author{Piotr Wielg{\'o}rski}
\affiliation{Nicolaus Copernicus Astronomical Center, Polish Academy of Sciences, Bartycka 18, 00-716 Warsaw, Poland}

\author{Grzegorz Pietrzy{\'n}ski}
\affiliation{Nicolaus Copernicus Astronomical Center, Polish Academy of Sciences, Bartycka 18, 00-716 Warsaw, Poland}
\affiliation{Universidad de Concepcion, Departamento de Astronomia, Casilla 160-C, Concepcion, Chile}

\author{Wolfgang Gieren}
\affiliation{Universidad de Concepcion, Departamento de Astronomia, Casilla 160-C, Concepcion, Chile}
\affiliation{Millenium Institute of Astrophysics, Santiago, Chile}

\author{Marek G{\'o}rski}
\affiliation{Universidad de Concepcion, Departamento de Astronomia, Casilla 160-C, Concepcion, Chile}
\affiliation{Millenium Institute of Astrophysics, Santiago, Chile}

\author{Rolf-Peter Kudritzki}
\affiliation{Institute for Astronomy, University of Hawaii at Manoa, 2680 Woodlawn 
Drive, Honolulu HI 96822, USA}

\author{Bart{\l{}}omiej Zgirski}
\affiliation{Nicolaus Copernicus Astronomical Center, Polish Academy of Sciences, Bartycka 18, 00-716 Warsaw, Poland}

\author{Fabio Bresolin}
\affiliation{Institute for Astronomy, University of Hawaii at Manoa, 2680 Woodlawn Drive, 
Honolulu HI 96822, USA}

\author{Jesper Storm}
\affiliation{Leibniz-Institut f\"ur Astrophysik Potsdam, An der Sternwarte 16, 14482, Potsdam, Germany}

\author{Noriyuki Matsunaga}
\affiliation{Department of Astronomy, The University of Tokyo, 7-3-1 Hongo, Bunkyo-ku, Tokyo 113-0033, Japan}

\author{Dariusz Graczyk}
\affiliation{Nicolaus Copernicus Astronomical Center, Polish Academy of Sciences, Bartycka 18, 00-716 Warsaw, Poland}
\affiliation{Universidad de Concepcion, Departamento de Astronomia, Casilla 160-C, Concepcion, Chile}
\affiliation{Millenium Institute of Astrophysics, Santiago, Chile}

\author{Igor Soszy{\'n}ski}
\affiliation{Warsaw University Observatory, Al. Ujazdowskie 4, PL-00-478 Warsaw, Poland}

\begin{abstract}

Using high-quality observed period-luminosity relations in both Magellanic Clouds in VIJHKs bands
and optical and near-infrared Wesenheit indices we determine the effect of metallicity on Cepheid
P-L relations by comparing the relative distance between LMC and SMC
as determined from the Cepheids to the distance difference between the Clouds which has been derived with very high accuracy
from late-type eclipsing binary systems. Within an uncertainty of $3\%$ which is
dominated by the uncertainty on the mean metallicity difference between the Cepheid populations
in LMC and SMC we find metallicity effects smaller than $2\%$ in all bands and in the Wesenheit indices, consistent with a
zero metallicity effect. This result is valid for the metallicity range from -0.35 dex to -0.75 dex corresponding to the mean [Fe/H] values
for classical Cepheids in LMC and SMC, respectively. Yet most Cepheids in galaxies beyond the Local Group and located
in the less crowded outer regions of these galaxies do fall into this metallicity regime, making our result important
for applications to determine the distances to spiral galaxies well beyond the Local Group.

Our result supports previous findings which indicated a very small metallicity effect on the near-infrared absolute magnitudes
of classical Cepheids, and resolves the dispute about the size and sign of the metallicity effect in the optical spectral range.
It also resolves one of the most pressing problems in the quest towards
a measurement of the Hubble constant with an accuracy of $1\%$ from the Cepheid-supernova Ia method.
\end{abstract}


%+++++++++++++introduction+++++++++++++++++

\section{Introduction}
Classical Cepheids are prime distance indicators and have been widely used to calibrate the first rungs of the
extragalactic distance scale using the famous period-luminosity (PL) relation (more recently also called the "Leavitt Law").
In the ongoing quest to achieve a determination of the Hubble constant with an accuracy of 1\%, Cepheid variables have been 
particularly useful 
to calibrate the peak luminosities of Type Ia supernovae which seem to offer the most promising route to achieve
such an accurate measurement of $H_0$ \citep[][]{2016ApJ...826...56R}. The two main limitations to further reduce the systematic uncertainty 
on $H_0$ as determined from the Cepheid-SN Ia method currently come from the 
uncertainty on the distance of the adopted fiducial galaxy to which the Cepheid PL relations observed in SN host galaxies 
are tied,
and from the ill-known effect metallicity might have on Cepheid absolute magnitudes, particularly in near- and 
mid-infrared bands
which are used in modern Cepheid distance work to reduce the effect of interstellar extinction on the results \citep[e.g.][]{2005ApJ...628..695G,2012ApJ...758...24F}.

While the situation with respect to the absolute distance of a suitable Cepheid fiducial galaxy has considerably improved
with the 2.2\% determination of the distance of the Large Magellanic Cloud (LMC) by \citet{2013Natur.495...76P} from
a sample of late-type eclipsing binaries, and the potential to push the accuracy of the LMC distance to even 1\% with
the same method by improving the surface brightness-color relation which serves to calculate the angular diameters
of the binary component stars \citep[e.g.][]{2017ApJ...837....7G}, the situation is much more confused with respect to the
knowledge of how exactly Cepheid absolute magnitudes depend on metallicity. A number of empirical tests have been made
over the years \citep[][]{1990ApJ...365..186F,1998ApJ...498..181K,2006ApJ...652.1133M,2008A&A...488..731R,2009MNRAS.396.1287S} which 
mostly favored the view that in optical bands metal-rich Cepheids are intrinsically brighter than
their metal-poor counterparts of the same pulsation period, but this result has been challenged by the work of
\citet{2008A&A...488..731R} who found an effect in the opposite sense. \citet{2011ApJ...734...46F} suggested that the
metallicity effect vanishes in the near-infrared (HK) domain while being significant at optical and mid-infrared
wavelengths, but with opposite signs. \citet{2011A&A...534A..95S}, using the Infrared Surface Brightness variant of the
Baade-Wesselink method \citep{1997A&A...320..799F,2005ApJ...627..224G} to determine the distances to Milky Way, LMC
and SMC
Cepheid samples, and comparing the resulting absolute PL relations of Milky Way and Magellanic Cloud Cepheids found a small effect
in all optical and near-infrared bands, but the uncertainty of this determination was in the order of $\pm 0.1$ mag/dex
which needs to be beaten down significantly in order to achieve a 1\% measurement of $H_0$. Theoretical work has been done,
among others, by \citet{2010ApJ...715..277B} but the uncertainties on the adopted input physics for the models are currently too large 
to allow to clearly support either of the empirical results. Clearly, a truly accurate empirical test is long overdue.

As stated above, a breakthrough in the century-long effort to accurately measure the distances to the Magellanic Clouds
has been made by using the very rare well-detached eclipsing binary systems being composed of two similar red giant stars for this
purpose. \citet{2013Natur.495...76P} have shown that the distance of one such system can be determined with an accuracy
close to 2\% if the data for the analysis (optical light curves, orbital radial velocity curve, and the out-of-eclipse
K-band magnitude of the system) are of very high precision. From eight systems, Pietrzynski et al. determined the LMC distance
with an accuracy of 2.2\%; the dominant source of systematic error in this process is the adopted surface brightness-color
relation which our group is currently improving by interferometric angular diameter measurements of a large number
of nearby red clump giants. The distance determined in this way is practically independent of metallicity and the assumed
reddening. Similarly, our group has succeeded to determine the distance to the central part of the
Small Magellanic Cloud (SMC) with an accuracy of 3\% using five late-type systems resembling those in the LMC used
by \citet{2013Natur.495...76P}, and using exactly the same prescripts in the analysis of these systems, and the distance
calculations \citep[][]{2014ApJ...780...59G}. As a consequence, the relative distance between LMC and SMC is an extremely well-determined number
from this work since the systematic errors cancel out, and the statistical uncertainties are in the order of 1\%
or less. 

In our long-term Araucaria Project, we have demonstrated how the combination of optical and near-infrared photometry
of Cepheids in combination with a reddening law can be used to very accurately measure the distance and mean reddening
of a target galaxy
with respect to the LMC \citep[e.g.][]{2005ApJ...628..695G,2006ApJ...642..216P,2006ApJ...648..375S}. In this paper, 
we are going to employ the very extensive optical and near-infrared photometric
datasets of Cepheids in both Clouds to construct PL relations (see section 2) and measure the relative distance between
SMC and LMC from the observed PL relations in optical (V,I) and near-infrared (J,H,Ks) photometric passbands. We can then compare
the distance difference between the Clouds measured from the Cepheids with the value obtained from the late-type eclipsing binaries, 
and interpret the
offset from the binary-based distance difference as due to the effect of metallicity on the Cepheid absolute magnitudes
in the corresponding band, assuming an average metallicity difference between the classical Cepheid populations
in LMC and SMC of 0.406 dex, the SMC Cepheids being more metal-poor by this amount (see section 5 of this paper).
Since the Cepheid PL relations in LMC and SMC are extremely well determined from large samples of stars with very accurate
photometry, the magnitude offsets, for a given band, between LMC and SMC can be measured very accurately, so we can
finally expect a measurement of the size, and sign of the metallicity effect in the VIJHKs bands which is approaching
the 1\% accuracy needed in the context of a precision measurement of $H_0$.

\section{Cepheid samples in the LMC and SMC}
In our work we used data from two surveys covering major parts of the Magellanic Clouds. The optical data come from the  
Optical Gravitational Lensing Experiment \citep[OGLE, e.g.][]{2015AcA....65....1U}. In the course of OGLE project, the Magellanic Clouds 
 have been observed since 1996 using the 1.3 meter telescope at the Las Campanas Observatory (Chile).
In particular during the third phase of this project (OGLE-III),  V and I band light curves of about 1700 classical Cepheids 
pulsating in the fundamental mode (FU) were obtained in the LMC 
\citep{2008AcA....58..163S}, and about 2000 FU Cepheids were observed in the SMC \citep{2010AcA....60...17S}. Multi-epoch photometry 
collected over 8-13 years was carefully transformed onto the 
system of Landolt  \citep{2008AcA....58..163S}.
From the OGLE-III catalog we adopted  magnitudes, coordinates and periods of the Magellanic Cloud Cepheids.
All stars brighter than about 12.5 mag are saturated on the OGLE images. This saturation limit  corresponds to Cepheids with periods of 
30 days. Therefore we complemented 
the LMC Cepheid sample with 28 bright FU Cepheids observed in the OGLE-III shallow survey \citep{2013AcA....63..159U}. This 
complementary survey was tailored for bright stars. All the data were collected with the same instrumental system 
and were tied to the OGLE III photometric system. As a result we have an extremely homogenous sample of optical light curves for all Cepheids 
in both Magellanic Clouds. 

The near-Infrared data in the J, H and Ks bands come from observations collected  with the 1.4 meter IRSF telescope located at the South African 
Astronomical Observatory Sutherland site \citep{2007PASJ...59..615K}. This catalog contains  single epoch observations. From these data we calculated mean 
magnitudes following the procedure described in \citet{2005PASP..117..823S}. We estimated mean magnitudes in all bands with 
a precision close to 0.03 mag. In order to transform the mean magnitudes of the Cepheids to the 2MASS system we applied the formulae given 
in \citet{2007PASJ...59..615K}.

Due to the observed nonlinearity of the Cepheid period-luminosity relation in the SMC at the very short-period end \citep{1999A&A...348..175E}
we rejected Cepheids pulsating with periods shorter than 2.5 days in this galaxy. In this way we ended up with samples
of nearly 1800 LMC and about 900 SMC fundamental mode Cepheids, each Cepheid in these samples having accurate mean 
magnitudes  in 5 filters (VIJHKs). Moreover, we calculated Wesenheit indices for combinations of V, I and J, Ks filters. 
The Wesenheit index for a set of 
two filters (X and Y)  is given by the following formula: $W_{XY}=m_Y-\frac{A_Y}{E(X-Y)}(m_X-m_Y)$. Such combinations 
of magnitudes for V, I and J, Ks filters are widely used for distance determination since they do not depend on reddening 
\citep[if assumed reddening law is correct, e.g.][]{1985ApJ...298..340M}. We adopted extinction coefficients for our filters using the \citet{1994ApJ...422..158O} and \citet{1989ApJ...345..245C} formulas, which lead to
$\frac{A_I}{E(V-I)}=1.55$ and $\frac{A_{Ks}}{E(J-Ks)}=0.69$.

\section{Relative distance between the SMC and LMC from classical Cepheids}

\subsection{Cepheid P-L relations in the Magellanic Clouds}

Applying the method of least squares and $3\sigma$ clipping we fitted straight lines $m=a\log P +b$ to obtain the coefficients of the observed
period-luminosity (P-L) relations for the LMC Cepheids. Results are presented in Table \ref{tab_lmc}. Data and fitted 
relations are also shown in  Figure 1.
 To check the agreement of the slopes of P-L relations 
 in LMC and SMC we also  obtained free fits to the SMC data, with results also given in Table 1 and fits plotted
in Figure 2. As can be appreciated the 
 agreement between the slope  measured in the LMC and SMC is very good, close to the combined  $1\sigma$ in all bands except V, where
 the discrepancy is close to the combined $3\sigma$ uncertainty on the slopes. 
 Since the slopes of the LMC P-L relations are determined with higher accuracy than those of the corresponding SMC relations,
 we adopted them for the SMC. Forcing the LMC slopes to the SMC P-L relations leads to the zero points given in Table 2 which we used
 to measure the observed SMC distance moduli in the different photometric bands. As previously, we used 
$3\sigma$ clipping to remove outstanding points. The last column of Table 2 gives the relative distance moduli
between the LMC and SMC obtained in this process for the different bands. In the case of the Wesenheit indices, these values are 
already the true, reddening-free relative distance  moduli, while in the case of the other bands the values are reddened, e.g. not corrected
for extinction. Our results are generally in very good agreement with the results obtained by other authors 
based on similar data for Cepheids in the Magellanic Clouds \citep[e.g.][]{2004AJ....128.2239P,2012MNRAS.424.1807R,2013ApJ...764...84I,2015AJ....149..117M,2015ApJ...808...67N,2016ApJS..224...21R}. 
Observed small differences are probably due to different samples of Cepheids  used by different authors. 

\subsection{Relative distance and reddening from multiband Cepheid P-L relations}

Based on the observed relative distance moduli obtained in the previous section and assuming the extinction law 
of \citet{1994ApJ...422..158O} and \citet{1989ApJ...345..245C} we will now measure
the true relative LMC-SMC distance modulus, and the relative reddening following our previous work in the Araucaria Project \citep[e.g.][]{2009ApJ...700.1141G,2006ApJ...642..216P}. The relative true distance modulus can be written as: \\
 
 \begin{equation}
(m-M)^{SMC}_{0} - (m-M)^{LMC}_{0} = \Delta(m-M)_{0}   =  \Delta (m-M)_{\lambda}  - \Delta E(B-V)*R_{\lambda} \\
\end{equation}
where $\lambda$ is a given filter,  $\Delta (m-M)_{\lambda}$ are the corresponding values from the last column of Table 2, $R_{\lambda}$ are the ratios 
of total to selective absorption 
 calculated from the \citet{1994ApJ...422..158O} and \citet{1989ApJ...345..245C} reddening law, and  $\Delta E(B-V) = E(B-V)_{SMC} - E(B-V)_{LMC}$ is the relative mean reddening 
 affecting the Cepheid samples in the LMC and SMC. 
 
Fitting a straight line to this relation with the data given in columns 2 and 3 of Table 3, we obtain the following results:

\begin{equation}
\Delta (m-M)_{0}=0.481\pm0.004  \mathrm{ mag},
\end{equation}
\begin{equation}
\Delta E(B-V)=-0.045\pm0.003  \mathrm{ mag}.
\end{equation}

The fit is shown in Fig. 3. Our result shows that the Cepheids in the SMC are on average by 0.045 mag less reddened than 
the LMC Cepheids.
In order to check on this result we used  the reddening maps of the LMC and SMC obtained by \citet{2011AJ....141..158H} and calculated 
the reddening for all studied 
Cepheids. From these maps which are based on red clump stars we find that our sample of Cepheids in the  SMC is 
on  average 0.036 mag less reddened 
than the sample in the LMC, in excellent agreement with our result from the multiband Cepheid  P-L relations in the present study.

\section{Relative distance between SMC and LMC from late-type eclipsing binaries}

Recently our group has measured precise and accurate geometrical distances  
to both Magellanic Clouds using late-type, fully detached eclipsing binary systems consisting of pairs of red giants
having typically 3-4 solar masses, very similar to the masses of short-period classical Cepheids. From 8 such systems, \citet{2013Natur.495...76P} 
obtained an LMC distance modulus of 18.493 $\pm$  0.008 (stat) $\pm$ 0.048 (sys) mag. In \citet{2013Natur.495...76P} it was demonstrated that the
effect of the geometrical depth of the LMC on this result is negligible.

From the analysis of 5 systems in the SMC resembling those in the LMC
we measured a distance modulus of 18.965 $\pm$ 0.025 (stat) $\pm$ 0.048 (sys) for the SMC barycenter \citep{2014ApJ...780...59G}.
As discussed in that paper, this value of the SMC distance should be only marginally affected by the more complicated,
as compared to the LMC, geometrical structure of the SMC.

 We recall that the distances of these systems were measured using the simple equation:
$ d(pc) = 1.337 \times 10^{-5}r(km)/ \varphi(mas)$.
The linear diameter (r) comes from the analysis of the system while 
the angular diameter ($\varphi$) is derived from a surface brightness-color 
relation (e.g.  $ m_{0} = S - 5 log(\varphi)$, where S is
the surface brightness in a given band, and $ m_{0}$ is the unreddened magnitude 
of a given star in this band).  The surface brightness - color relation (SBCR) using the $(V-K)$ color is very well established 
(to about 2 \%) from interferometric observations \citep[e.g.][]{2005MNRAS.357..174D,2004A&A...428..587K}. 
This method depends only very weakly on reddening and the adopted reddening law. It is also practically independent of metallicity 
\citep[][]{2013Natur.495...76P,2001AJ....121.3089T}. A change of 1 dex would produce a change of Sv of about 0.006 mag, which corresponds
to a 0.3 \% change in the distance determination. Since our samples of eclipsing systems in both Clouds differ in metallicity
by 0.4 dex (see section 5) this could introduce an error at the level of 0.1\% only, therefore we will neglect it. 

The systematic uncertainty on our distance measurement of LMC and SMC from late-type eclipsing binaries comes from the calibration 
of the SBCR, and the accuracy of the zero points in our photometry. Since the same photometric system and the same SBCR was used to calculate 
the distances to the LMC and SMC, the {\it relative} distance is dominated only by statistical errors and it equals to:\\

\begin{equation}
\Delta (m-M)_{0}^{ecl} = 0.472 \pm 0.026 \space\mathrm{mag}
\end{equation}

We emphasize that the studied  eclipsing binaries are composed of stars which are in the same 
evolutionary phase as Cepheids (e.g. helium-burning giants), which should imply that their spatial distributions
in the Clouds are very similar to those of the classical Cepheids. Therefore we do not expect any significant 
influence of the LMC and SMC depth effects on our results.

\section{Metallicities of the young stellar populations in the Magellanic Clouds}

We have surveyed the literature for metallicity determinations of young stars in the Magellanic Clouds focussing on studies,
which employ exactly the same technique in their analyses of both Clouds. In this way we intend to obtain abundance differences 
between the two Clouds which are as little as possible influenced by any systematic effects caused by the analysis method and the
atmospheric structures of stars of different type. \citet{1995A&A...293..347H} and \citet{1999A&A...} obtained a [Fe/H] difference between LMC
and SMC of -0.485 $\pm$ 0.12 dex from F and K supergiants. From classical Cepheids \citet{1998AJ....115..605L} determined a difference
of -0.38 $\pm$ 0.13 dex, while \citet{2008A&A...488..731R} found a value of -0.42 $\pm$ 0.15 dex. All three studies were carried out
in LTE. From \citet{2015ApJ...806...21D} we obtained a logarithmic metallicity difference [Z] of -0.17 $\pm$ 0.12 dex from their NLTE study
of red supergiant stars, which included the elements Fe, Ti, Si, and Mg (note that we have excluded two outliers in the SMC and one
in the LMC from their sample).

We have also included two investigations of hotter stars. From \citet{2007A&A...471..625T} we infer a difference in metallicity [Z]
of -0.36 $\pm$ 0.04 dex including the elements O, Mg, Si, and Fe. Note that the former three elements were treated in NLTE whereas
the analysis of Fe is based on LTE. \citet{2017ApJ...} and \citet{2010thesis} in their study of A supergiant stars used the same set
of NLTE radiative transfer calculations for Fe and Mg in the LMC and SMC, respectively, to obtain a metallicity difference
of -0.385 $\pm$ 0.16 dex.

The mean metallicity difference between SMC and LMC obtained from these six studies is -0.367 $\pm$ 0.106 dex. With respect to this 
mean value the Davies et al. difference is a 2-$\sigma$ outlier. Excluding the Davies et al. value we obtain a mean difference
of $-$0.406 with a standard deviation of $\sigma$ = 0.048 dex. In the following, we will use this value.

\section{Effect of metallicity on Cepheid absolute magnitudes in the optical and near-infrared}

We adopt the relative distance moduli obtained for the VIJHKs, and the two Wesenheit bands (last column of Table 2) and correct them 
for the relative reddening (0.045 mag) measured from the analysis in section 3.2. This yields the values
quoted in column 4 of Table 3.  Assuming that the differences of the values of the unreddened relative 
distances obtained from the Cepheids on the one hand, and eclipsing binaries on the other hand are due to the $-$0.406 dex difference 
in mean metal content 
of the Cepheids in LMC and SMC, we calculate the metallicity effect $\gamma$ in each filter in the following way: 

\begin{equation}
\gamma = \frac{ \Delta (m-M)_{0}^{cep} - \Delta (m-M)_{0}^{ecl} } {  \Delta [Fe/H] }
\end{equation}

The values of $\gamma$ for the different filters and their uncertainties are given in the last column of Table 3.

Statistical tests have revealed slight nonlinearities near 10 days in the JHK P-L relations in the LMC
\citep[e.g.][]{2016MNRAS.457.1644B}. 
In order to test the stability of the derived $\gamma$ values with regard to the chosen samples of LMC and SMC Cepheids, we repeated the analysis
a) introducing the same period cutoff at 2.5 days for the LMC Cepheid sample as done for the SMC sample; b) retaining in both Clouds
only Cepheids with periods longer then 10 days; and c) comparing the magnitudes derived from the free fits to the full Cepheid samples at a specific
intermediate period of 10 days. In Table 4, we give the results for the relative true distance between SMC and LMC for these cases,
as well as the reddening differences, and the resulting metallicity effects together with their uncertainties. As can be appreciated
from the numbers in Table 4, in case a) the changes are totally negligible; in case b) the derived metallicity effect is smaller than 1\% for all bands
except V and I where it is about 1.5\%; and in case c) the metallicity effect is smaller than 2\% in all bands. Most importantly, in all
these cases the metallicity effect is consistent with a zero effect in all bands, within the very small uncertainties. We conclude that
the details of the selection of the Cepheid samples in LMC and SMC do not have any significant effect on the values for the metallicity effect
we obtain in the considered photometric bandpasses.

\section{Discussion}

As can be seen from the values in Table 3, our study reveals a very small metallicity effect, less than 2\% in all bands investigated in this study.
In particular, the effect is less than 1\% in the optical I and near-infrared H and Ks bands, as well as in the two Wesenheit indices.
Within the small uncertainty of 3\% of the $\gamma$ values in Table 3, the metallicity effect derived in this paper is clearly consistent
with a null effect in all optical and near-infrared bands; the same holds for the Wesenheit indices. The same conclusion was reached earlier by 
\citet{2011ApJ...741L..36M} for the optical VI Wesenheit index. Obviously, this conclusion of a vanishing metallicity effect is restricted
to the metallicity range of $-$0.35 dex $>$ [Fe/H] $>$ $-$0.75 dex comprised by the classical Cepheid populations in LMC and SMC.

The sign of the small metallicity effect we find in all bands is in agreement with the study of \citet{2008A&A...488..731R}, and at odds
with all the studies which derived the metallicity effect by comparing inner- and outer-field
Cepheid samples in spiral galaxies and assuming a value of the metallicity gradient in the disc of these galaxies (usually determined
from H II region oxygen abundances), a method which was
pioneered by \citet{1990ApJ...365..186F} by a comparison of three fields at different galactocentric distances in the
Andromeda galaxy. The most likely cause that these studies consistently found the more metal-rich Cepheids located in the inner 
galaxy regions to be significantly brighter than their more metal-poor counterparts in the outer part of the galaxy is the effect of blending
which affects the crowded inner regions more severely than the outer regions of a spiral galaxy, even at HST resolution
as for example in the study of M 101 Cepheids by \citet{1998ApJ...498..181K}. 

We note that using gas-phase oxygen abundances in place of the stellar metallicities we presented in Sect. 5 does not
change our conclusions. Taking the compilation of O/H abundances for HII regions in the LMC and the SMC by \citet{2011ApJ...729.56B}, who consistently derived them from the calculation of the gas electron temperature, we obtain a difference
of -0.33 $\pm$ 0.13 dex between the SMC and the LMC. This difference yields a $\gamma$ value that is fully compatible
with the one based on stellar metallicities alone.

Regarding the near-infrared spectral region, our current result is in line with the results of \citet{2011ApJ...734...46F}, and of \citet{2011A&A...534A..95S}
who also found with different methods a vanishing metallicity effect in the near-infrared regime, particularly in the H and K bands. The results
of \citet{2011A&A...534A..95S} obtained from an application of the Near-Infrared Surface Brightness Technique to large samples of Milky Way and LMC,
and a small sample of SMC
classical Cepheids are confirmed and strenghtened by the extension of this work to a much larger sample of SMC Cepheids \citep[][]{2017A&A...}. The present
determination however is more accurate, by a factor of at least 2, than these previous measurements. Regarding the optical regime,
the cited previous studies did find significant metallicity effects, particularly in the V band, whereas our current study
indicates that the effect in V and I is at most 2\% and consistent with a null effect in these bands, too. 

The main assumption underlying our study is the universality of the slope of the Cepheid P-L relation in all bands we have considered,
at least in the range from $-$0.35 to $-$0.75 dex corresponding to the Cepheid metallicities in LMC and SMC.
While there have been a few studies challenging this assumption \citep[e.g.][]{2008ApJ...686..779S}, most empirical evidence, particularly from
studies of Cepheids in nearby Local Group galaxies with excellent datasets as in our own Araucaria Project have confirmed that the observed slope 
of the P-L relation is, within small uncertainties, identical for the Cepheid populations in a number of galaxies having a range of metallicities
of their young stellar populations, from LMC abundances down to about -1.0 dex in the dwarf irregular galaxies WLM and IC 1613
\citep[][]{2006ApJ...648.1007B,2007ApJ...671.2028B}.  This seems to be particularly true for the near-infrared JHK bands. For example, \citet{2006ApJ...642..216P}, in their distance study of the very  metal-poor Local Group galaxy IC 1613 found excellent agreement between the observed slopes of the P-L relations in the J and K bands and those of the corresponding LMC Cepheid relations of \citet{2004AJ....128.2239P}. Claims that P-L relation slopes in different
optical bands might vary with metallicity \citep[e.g.][]{2010ApJ...715..277B} were almost exclusively based on Cepheid data from external galaxies
well beyond the Local Group where crowding and blending of the Cepheids is likely to be a problem which affects the Cepheid
photometry at an unknown level. Such blending will affect the Cepheid photometry in a differential way, 
affecting the short-period fainter Cepheid more strongly than the brighter long-period Cepheids, which could introduce spurious changes
in the slopes of the observed P-L relations. It is certainly much safer to rely on Local Group Cepheid data and observed P-L relations, where the
best Cepheid photometric data are reasonably unaffected by the very complicated blending and crowding problem.

\section{Conclusions}

We have used the very high-quality optical photometric data of the OGLE Project in the V and I bands together with near-infrared JHKs
photometry obtained on the IRSF telescope in South Africa to establish very precise period-luminosity relations in both Magellanic Clouds.
These data have been used to determine the effect of metallicity on Cepheid absolute magnitudes in these bands,
and on the optical and near-infrared Wesenheit indices. We did this by comparing the relative distance between LMC and SMC as determined
from the Cepheids in the different photometric bands to the distance difference between the Clouds which has been measured very accurately
with late-type eclipsing binaries in both Clouds. The comparison yields a metallicity effect smaller than 2\% in all bands and in the
Wesenheit indices, which within the 3\% uncertainty of the determination means that there is no significant metallicity effect in neither
of the bands.

We selected for our study Cepheids in the central parts of the Magellanic Clouds where observed depth effects are small. 
Since the components of our binary systems (helium burning giants) are expected to have the same spatial distribution as the Cepheids and 
we have been able to use sizeable samples  
of eclipsing binaries in both Clouds, our results should not be affected in any significant way by the geometrical extension of both Magellanic Clouds \citep{2013Natur.495...76P,2014ApJ...780...59G}. Reported differences between P-L relations based on different samples of Cepheids 
 \citep[e.g.][]{2004AJ....128.2239P,2012MNRAS.424.1807R,2013ApJ...764...84I,2015AJ....149..117M,2015ApJ...808...67N,2016ApJS..224...21R} are very small and confirm that our results are not 
 significantly affected by the depth effect.

While the present result is restricted to the metallicity range comprised by the Cepheids in both Magellanic Clouds, it is this the metallicity
range into which the Cepheids observed in distant spiral galaxies at several Mpc typically fall. While in massive spiral galaxies Cepheid variables 
very close to the centers may have abundances even above solar, it is the Cepheids at relatively large galactocentric distances in these galaxies
which are typically used for determining the distances of their host galaxies because they are better resolved and less affected by crowding
in the images. Therefore the metallicity range from LMC to SMC abundances seems to be the most relevant range for Cepheid distance work to spiral galaxies
located at distances of about 5-20 Mpc, which makes the present results for the metallicity effect on Cepheids relevant and important for
the determination of Cepheid distances to SN Ia host galaxies in the near-infrared (H-band) regime using HST or, in the near future, the JWST
space telescope.

\acknowledgements
The research leading to these results has received funding from the European Research Council (ERC) under 
the European Union's Horizon 2020 research and innovation program (grant agreement No 695099).
WG, MG and DG gratefully acknowledge financial support for this work from  the Millenium Institute of Astrophysics (MAS) 
of the Iniciativa Cientifica Milenio del Ministerio de Economia, Fomento y Turismo de Chile, project IC120009.
We (WG, GP and DG) also very gratefully acknowledge financial support for this work from the BASAL Centro de Astrofisica
y Tecnologias Afines (CATA) PFB-06/2007. We also acknowledge support from the Polish National Science Center
grant MAESTRO DEC-2012/06/A/ST9/00269. Based on observations made with ESO telescopes under programme 098.D-0263(A,B), 097.D-0400(A), 097.D-0150(A), 097.D-0151(A) and CNTAC programme CN2016B-38, CN2016A-22, CN2015B-2, CN2015A-18. Last not least, we are grateful to the OGLE and IRSF team members for providing data of outstanding quality  which made 
this investigation possible.

%++++++++++++++bibliography++++++++++++++++++

\clearpage

%-------tables--------------------
%relations in LMC and SMC 
\begin{table}[ht]
\begin{center}
\caption{Free linear fits to the observed  P-L relations in the LMC and SMC.}
\label{tab_lmc}
\begin{tabular}{c c c c c}\\
\hline
\hline
Filter & $a$ & $b$ & $\sigma$ & $N$ \\
\hline
\multicolumn{5}{c}{LMC}\\
\hline
$V$ & $-2.779\pm0.021$ & $17.543\pm0.014$ & $0.228$ & $1794$\\
$I$ & $-2.977\pm0.015$ & $16.892\pm0.010$ & $0.154$ & $1769$\\
$J$ & $-3.118\pm0.011$ & $16.431\pm0.008$ & $0.120$ & $1740$\\
$H$ & $-3.224\pm0.009$ & $16.152\pm0.006$ & $0.097$ & $1743$\\
$Ks$ & $-3.247\pm0.009$ & $16.095\pm0.006$ & $0.089$ & $1747$\\
$W_I$ & $-3.332\pm0.008$ & $15.904\pm0.006$ & $0.083$ & $1778$\\
$W_{JKs}$ & $-3.334\pm0.008$ & $15.857\pm0.005$ & $0.084$ & $1743$\\%OK--done (WV?)
\hline
\multicolumn{5}{c}{SMC}\\
\hline
$V$ & $-2.644\pm0.036$ & $17.792\pm0.026$ & $0.282$ & $955$\\
$I$ & $-2.947\pm0.027$ & $17.264\pm0.020$ & $0.223$ & $963$\\
$J$ & $-3.087\pm0.023$ & $16.858\pm0.017$ & $0.185$ & $907$\\
$H$ & $-3.184\pm0.021$ & $16.577\pm0.016$ & $0.166$ & $906$\\
$Ks$ & $-3.206\pm0.021$ & $16.530\pm0.015$ & $0.160$ & $901$\\
$W_I$ & $-3.330\pm0.019$ & $16.385\pm0.014$ & $0.146$ & $946$\\
$W_{JKs}$ & $-3.311\pm0.020$ & $16.322\pm0.014$ & $0.150$ & $896$\\%OK--done (WV?)
\hline
\end{tabular}
\end{center}
\end{table}

%relations in SMC with slopes from LMC

\begin{table}[ht]
\begin{center}
\caption{Zero points of SMC Cepheid P-L relations with slopes adopted from the LMC}
\label{tab_dst}
\begin{tabular}{c c c c c}\\
\hline
\hline
Filter & $b_{SMC}$ & $\sigma$ & $N$ & $\Delta(m-M)$ \\
\hline
$V$ & $17.883\pm0.009$ & $0.282$ & $954$ & $0.340\pm0.017$\\
$I$ & $17.285\pm0.007$ & $0.222$ & $962$ & $0.393\pm0.012$\\
$J$ & $16.880\pm0.006$ & $0.186$ & $908$ & $0.449\pm0.010$\\
$H$ & $16.604\pm0.006$ & $0.166$ & $906$ & $0.452\pm0.009$\\
$Ks$ & $16.558\pm0.005$ & $0.161$ & $901$ & $0.463\pm0.008$\\
$W_{VI}$ & $16.386\pm0.005$ & $0.145$ & $946$ & $0.482\pm0.008$\\
$W_{JKs}$ & $16.338\pm0.005$ & $0.150$ & $896$ & $0.481\pm0.008$\\%OK--done
\hline
\end{tabular}
\end{center}
\end{table}

%metallicity effect

\begin{table}[ht]
\begin{center}
\caption{Metallicity effect on Cepheid P-L relations in VIJHK, $W_{VI}$ and $W_{JK}$ bands}
\label{tab_met_eff}
\begin{tabular}{c c c c c c}\\
\hline
\hline
Filter & $R_{\lambda}$ & $\Delta(m-M)$ & $\Delta(m-M)_{0}^{cep}$ & $\Delta(m-M)_{0}^{cep}-\Delta(m-M)_{0}^{ecl}$ & $\gamma$\\
\hline
$V$ & $ 3.134$ & $   0.340 \pm  0.017$ & $   0.481 \pm 0.017$ & $0.009 \pm 0.031$ & $-0.022 \pm 0.076$\\ 
$I$ & $ 1.894$ & $   0.393 \pm  0.012$ & $   0.478 \pm 0.012$ & $0.006 \pm 0.029$ & $-0.015 \pm 0.071$\\ 
$J$ & $ 0.892$ & $   0.449 \pm  0.010$ & $   0.489 \pm 0.010$ & $0.017 \pm 0.028$ & $-0.042 \pm 0.069$\\ 
$H$ & $ 0.553$ & $   0.452 \pm  0.009$ & $   0.477 \pm 0.009$ & $0.005 \pm 0.028$ & $-0.012 \pm 0.069$\\ 
$Ks$ & $ 0.363$ & $   0.463 \pm  0.008$ & $   0.479 \pm 0.009$ & $0.007 \pm 0.028$ & $-0.017 \pm 0.069$\\%OK--calc gamma
$W_{VI}$ & - & - & $0.482\pm0.008$ & $0.010 \pm 0.027$ & $-0.025 \pm 0.067$\\
$W_{JKs}$ & - & - & $0.481\pm0.008$ & $0.009 \pm 0.027$ & $-0.022 \pm 0.067$\\
\hline
\end{tabular}
\end{center}
\end{table}

% NEW TABLE
%metallicity effect and relation nonlinearity                                                                                               

\begin{table}[ht]
\begin{center}
\caption{Dependence of the metallicity effect on chosen sample cutoffs}
\label{tab_met_eff_period}
\begin{tabular}{c c c c c c c c c c}
\hline
\hline
Periods in LMC, SMC & $\Delta(m-M)_0$ & $\Delta E(B-V)$ & $\gamma_V$ & $\gamma_I$ & $\gamma_J$ & $\gamma_H$ & $\gamma_{Ks}$ & $\gamma_{W_{VI}}$ & $\gamma_{W_{JKs}}$\\
\hline
$(0:100)$,$(2.5:100)$ & $0.480 \pm 0.004$ & $-0.045 \pm 0.003$ & $-0.022$ & $-0.015$ & $-0.042$ & $-0.012$ & $-0.017$ & $-0.025$ & $-0.022$\\
$(2.5:100)$,$(2.5:100)$ & $0.480 \pm 0.003$ & $-0.045 \pm 0.003$ & $-0.015$ & $-0.012$ & $-0.039$ & $-0.012$ & $-0.017$ & $-0.022$ & $-0.025$\\
both $>$ 10d& $0.472 \pm 0.005$ & $-0.037 \pm 0.005$ & $-0.029$ & $0.030$ & $-0.022$ & $0.007$ & $-0.010$ & $-0.010$ & $-0.017$\\
free fits mags at 10d & $0.478 \pm 0.010$ & $-0.033 \pm 0.008$ & $-0.037$ & $0.017$ & $-0.037$ & $0.022$ & $-0.039$ & $-0.027$ & $-0.039 $\\
\hline
\end{tabular}
\end{center}
\end{table}

\clearpage
\begin{figure}
\begin{center}
\includegraphics[width=12cm,angle=-90]{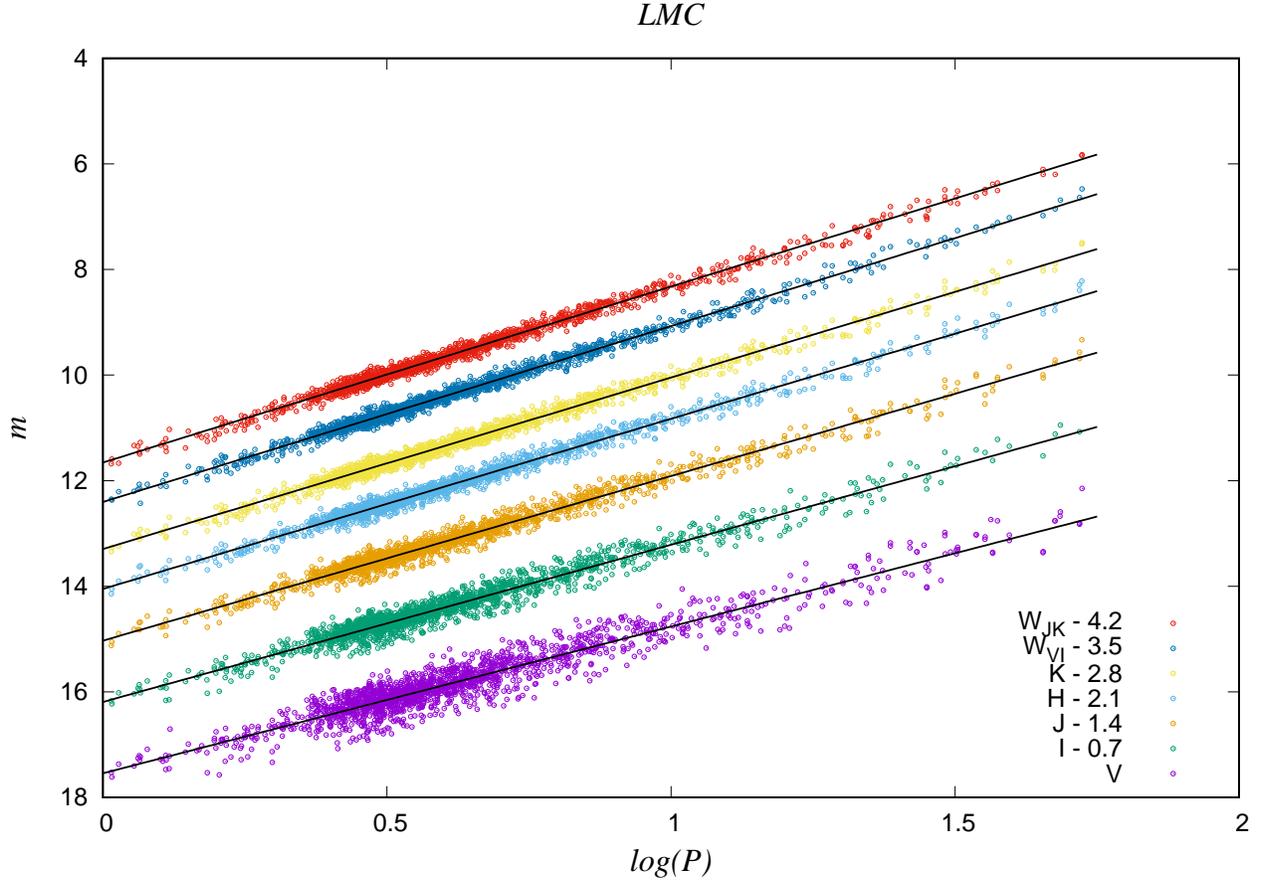}
\caption{Data and fitted P-L relations for $V$, $I$, $J$, $H$, $Ks$ filters, and $W_{VI}$ and $W_{JKs}$ Wesenheit indices  for the LMC Cepheids. The parameters of the fits are given in Table 1.}
\end{center}
\end{figure}

%relations in J,H,K,WJK
\begin{figure}
\begin{center}
\includegraphics[width=12cm,angle=-90]{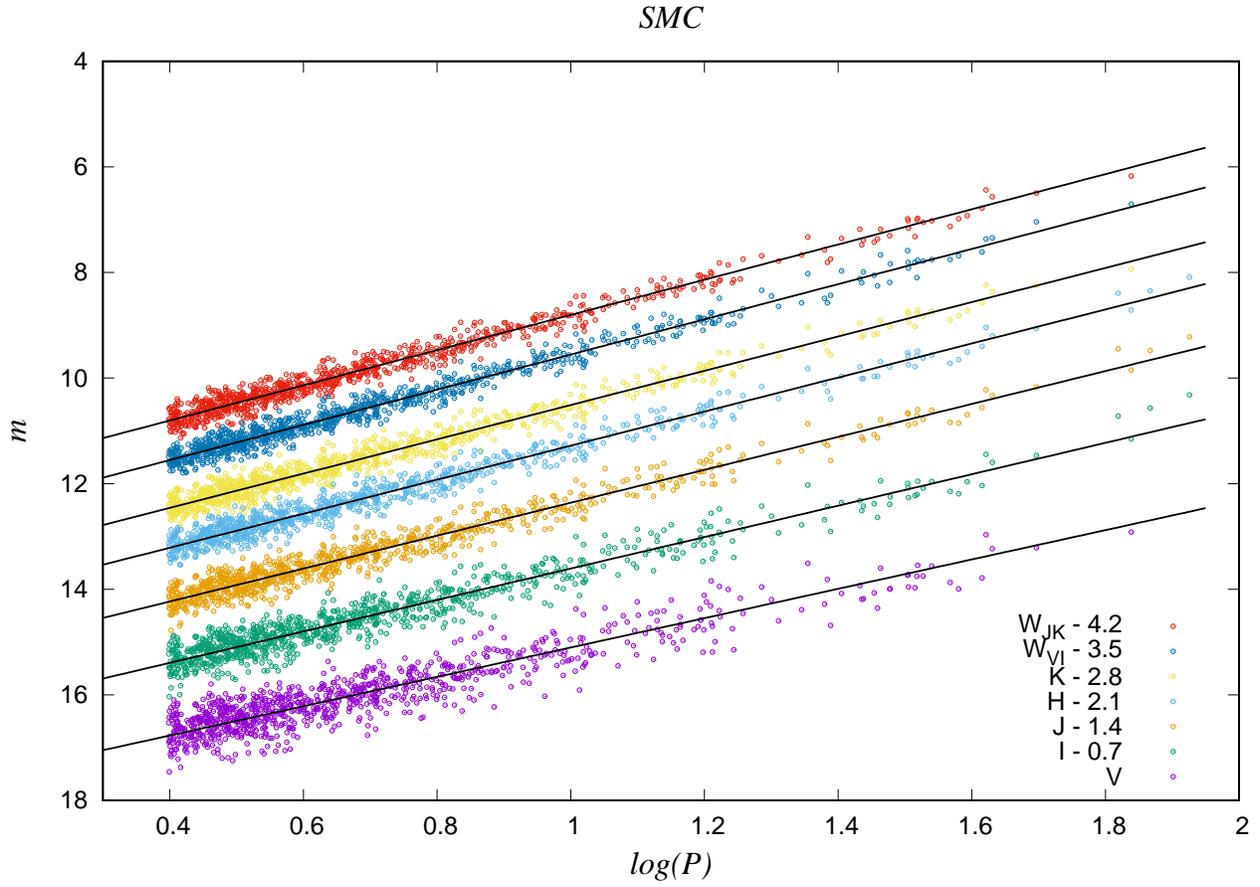}
\caption{
P-L relations for $V$, $I$, $J$, $H$, $Ks$ filters, and $W_{VI}$ and $W_{JKs}$ Wesenheit indices  for the SMC Cepheids.
Fits to a straight line are plotted, with the slopes adopted from the corresponding  LMC Cepheid relation. 
The zero points and dispersions of the P-L relations are given in Table 2.
}
\end{center}
\end{figure}

%main figure
\begin{figure}
\begin{center}
\includegraphics[width=12cm,angle=-90]{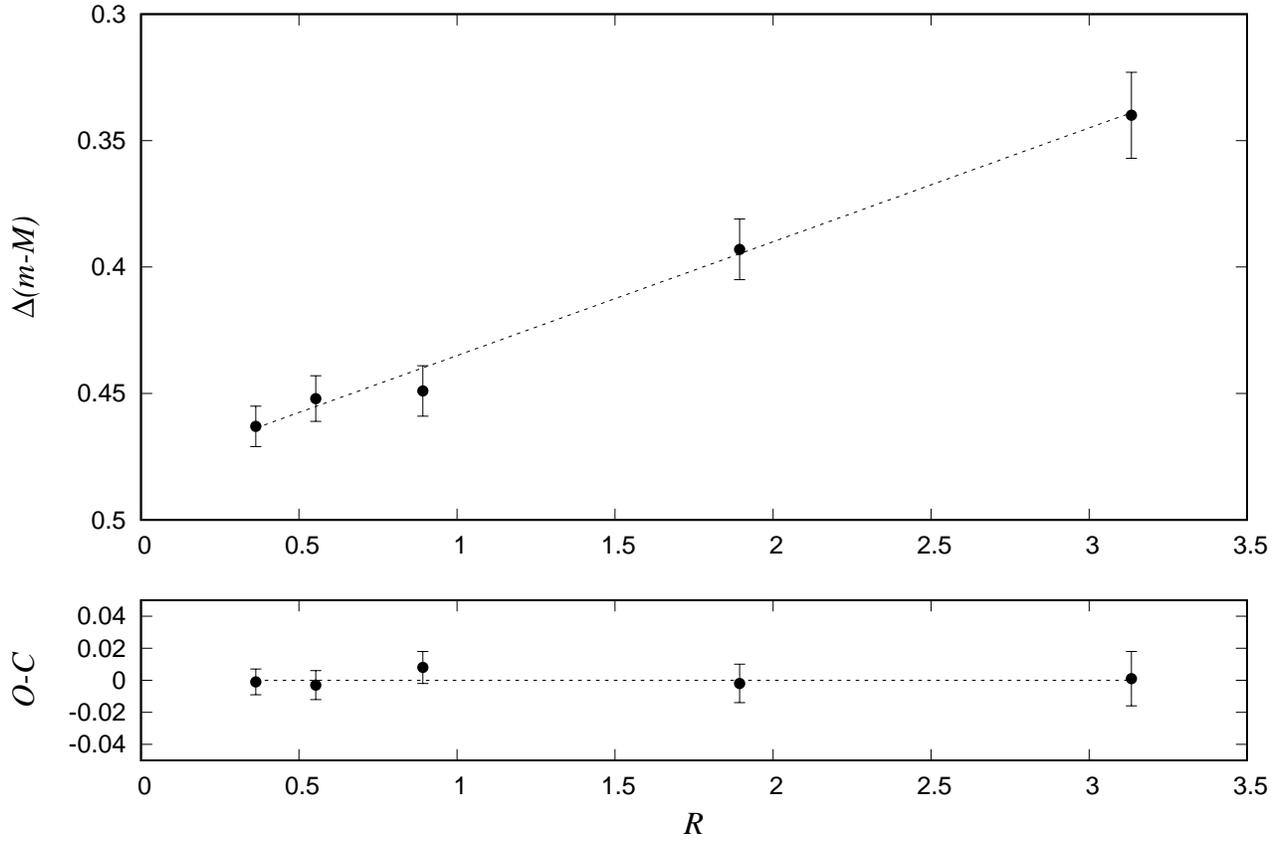}
\caption{Difference of the reddened distance moduli to the Magellanic Clouds determined in different photometric bands, 
plotted against the ratio of total to selective absorption. The slope of the linear fit  to
these data gives the difference of the amount of mean reddening affecting the Cepheids in LMC and SMC, and the
intersection of the best-fitting line yields the true, unreddened  difference of the distance moduli of the Clouds.
The lower panel shows the very small residuals of the data from the fitted line. The numerical results are given in
section 3.2.}
\end{center}
\end{figure}

\clearpage


\begin{thebibliography}{}

\bibitem[Bauer et al.(1999)]{1999A&A...348..175E} EROS Collaboration, Bauer, F., Afonso, C., et al.\ 1999, \aap, 348, 175 

\bibitem[Di Benedetto(2005)]{2005MNRAS.357..174D} Di Benedetto, G.~P.\ 2005, \mnras, 357, 174 

\bibitem[Bhardwaj et al.(2016)]{2016MNRAS.457.1644B} Bhardwaj, A., Kanbur, S.~M., Macri, S.M., Singh, H.~P., Ngeow, C.-C. \& Ishida, E.E.O.\ 2016, \mnras, 457, 1644

\bibitem[Bono et al.(2010)]{2010ApJ...715..277B} Bono, G., Caputo, F., Marconi, M., \& Musella, I.\ 2010, \apj, 715, 277 

\bibitem[Bresolin et al.(2006)]{2006ApJ...648.1007B} Bresolin, F., Pietrzy{\'n}ski, G., Urbaneja, M.~A., et al.\ 2006, \apj, 648, 1007 

\bibitem[Bresolin et al.(2007)]{2007ApJ...671.2028B} Bresolin, F., Urbaneja, M.~A., Gieren, W., Pietrzy{\'n}ski, G., \& Kudritzki, R.-P.\ 2007, \apj, 671, 2028

\bibitem[Bresolin(2011)]{2011ApJ...729.56B} Bresolin, F.\ 2011, \apj, 729, 56

\bibitem[Cardelli et al.(1989)]{1989ApJ...345..245C} Cardelli, J.~A., Clayton, G.~C., \& Mathis, J.~S.\ 1989, \apj, 345, 245 

\bibitem[Davies et al.(2015)]{2015ApJ...806...21D} Davies, B., Kudritzki, R.-P., Gazak, Z., et al.\ 2015, \apj, 806, 21 

\bibitem[O'Donnell(1994)]{1994ApJ...422..158O} O'Donnell, J.~E.\ 1994, \apj, 422, 158 

\bibitem[Fouque \& Gieren(1997)]{1997A&A...320..799F} Fouque, P., \& Gieren, W.~P.\ 1997, \aap, 320, 799 

\bibitem[Freedman \& Madore(1990)]{1990ApJ...365..186F} Freedman, W.~L., \& Madore, B.~F.\ 1990, \apj, 365, 186 

\bibitem[Freedman \& Madore(2011)]{2011ApJ...734...46F} Freedman, W.~L., \& Madore, B.~F.\ 2011, \apj, 734, 46 

\bibitem[Freedman et al.(2012)]{2012ApJ...758...24F} Freedman, W.~L., Madore, B.~F., Scowcroft, V., et al.\ 2012, \apj, 758, 24

\bibitem[Gieren et al.(2005a)]{2005ApJ...628..695G} Gieren, W., Pietrzy{\'n}ski, G., Soszy{\'n}ski, I., et al.\ 2005a, \apj, 628, 695 

\bibitem[Gieren et al.(2005b)]{2005ApJ...627..224G} Gieren, W., Storm, J., Barnes, T.~G., III, et al.\ 2005b, \apj, 627, 224  

\bibitem[Gieren et al.(2009)]{2009ApJ...700.1141G} Gieren, W., Pietrzy{\'n}ski, G., Soszy{\'n}ski, I., et al.\ 2009, \apj, 700, 1141 

\bibitem[Gieren et al.(2013)]{2013ApJ...773...69G} Gieren, W., G{\'o}rski, M., Pietrzy{\'n}ski, G., et al.\ 2013, \apj, 773, 69 

\bibitem[Gieren et al.(2017)]{2017A&A...} Gieren, W., Storm, J., Konorski., et al.\ 2017, \aap, in preparation

\bibitem[Graczyk et al.(2014)]{2014ApJ...780...59G} Graczyk, D., 
Pietrzy{\'n}ski, G., Thompson, I.~B., et al.\ 2014, \apj, 780, 59 

\bibitem[Graczyk et al.(2017)]{2017ApJ...837....7G} Graczyk, D., Konorski, P., Pietrzy{\'n}ski, G., et al.\ 2017, \apj, 837, 7

\bibitem[Haschke et al.(2011)]{2011AJ....141..158H} Haschke, R., Grebel, E.~K., \& Duffau, S.\ 2011, \aj, 141, 158

\bibitem[Hill et al.(1995)]{1995A&A...293..347H} Hill, V., Andrievsky, S., \& Spite, M.\ 1995, \aap, 293, 347 

\bibitem[Hill (1999)]{1999A&A...} Hill, V. \ 1999, \aap, 345, 430

\bibitem[Inno et al.(2013)]{2013ApJ...764...84I} Inno, L., Matsunaga, N., Bono, G., et al.\ 2013, \apj, 764, 84 

\bibitem[Kato et al.(2007)]{2007PASJ...59..615K} Kato, D., Nagashima, C., 
Nagayama, T., et al.\ 2007, \pasj, 59, 615 

\bibitem[Kennicutt et al.(1998)]{1998ApJ...498..181K} Kennicutt, R.~C., Jr., Stetson, P.~B., Saha, A., et al.\ 1998, \apj, 498, 181 

\bibitem[Kervella et al.(2004)]{2004A&A...428..587K} Kervella, P., Bersier, D., Mourard, D., et al.\ 2004, \aap, 428, 587 

\bibitem[Luck et al.(1998)]{1998AJ....115..605L} Luck, R.~E., Moffett, T.~J., Barnes, T.~G., III, \& Gieren, W.~P.\ 1998, \aj, 115, 605

\bibitem[Macri et al.(2006)]{2006ApJ...652.1133M} Macri, L.~M., Stanek, K.~Z., Bersier, D., Greenhill, L.~J., \& Reid, M.~J.\ 2006, \apj, 652, 1133

\bibitem[Macri et al.(2015)]{2015AJ....149..117M} Macri, L.~M., Ngeow, C.-C., Kanbur, S.~M., Mahzooni, S., \& Smitka, M.~T.\ 2015, \aj, 149, 117 

\bibitem[Madore(1985)]{1985ApJ...298..340M} Madore, B.~F.\ 1985, \apj, 298, 340

\bibitem[Majaess et al.(2011)]{2011ApJ...741L..36M} Majaess, D., Turner, D., \& Gieren, W.\ 2011, \apjl, 741, L36 

\bibitem[Ngeow et al.(2015)]{2015ApJ...808...67N} Ngeow, C.-C., Kanbur, S.~M., Bhardwaj, A., \& Singh, H.~P.\ 2015, \apj, 808, 67

\bibitem[Persson et al.(2004)]{2004AJ....128.2239P} Persson, S.~E., Madore, B.~F., Krzemi{\'n}ski, W., et al.\ 2004, \aj, 128, 2239 

\bibitem[Pietrzy{\'n}ski et al.(2006)]{2006ApJ...642..216P} Pietrzy{\'n}ski, G., Gieren, W., Soszy{\'n}ski, I., et al.\ 2006, \apj, 642, 216 

\bibitem[Pietrzy{\'n}ski et al.(2013)]{2013Natur.495...76P} Pietrzy{\'n}ski, G., Graczyk, D., Gieren, W., et al.\ 2013, \nat, 495, 76

\bibitem[Riess et al.(2016)]{2016ApJ...826...56R} Riess, A.~G., Macri, L.~M., Hoffmann, S.~L., et al.\ 2016, \apj, 826, 56

\bibitem[Ripepi et al.(2012)]{2012MNRAS.424.1807R} Ripepi, V., Moretti, M.~I., Marconi, M., et al.\ 2012, \mnras, 424, 1807 

\bibitem[Ripepi et al.(2016)]{2016ApJS..224...21R} Ripepi, V., Marconi, M., Moretti, M.~I., et al.\ 2016, \apjs, 224, 21 

\bibitem[Romaniello et al.(2008)]{2008A&A...488..731R} Romaniello, M., Primas, F., Mottini, M., et al.\ 2008, \aap, 488, 731

\bibitem[Sandage \& Tammann(2008)]{2008ApJ...686..779S} Sandage, A., \& Tammann, G.~A.\ 2008, \apj, 686, 779  

\bibitem[Schiller(2010)]{2010thesis} Schiller, F.\ 2010, PhD thesis, Naturwissenschaftliche Fakultaet, Friedrich-Alexander-Universitaet Erlangen-Nuernberg

\bibitem[Scowcroft et al.(2009)]{2009MNRAS.396.1287S} Scowcroft, V., Bersier, D., Mould, J.~R., \& Wood, P.~R.\ 2009, \mnras, 396, 1287 

\bibitem[Soszy{\'n}ski et al.(2005)]{2005PASP..117..823S} Soszy{\'n}ski, I., Gieren, W., \& Pietrzy{\'n}ski, G.\ 2005, \pasp, 117, 823

\bibitem[Soszy{\'n}ski et al.(2006)]{2006ApJ...648..375S} Soszy{\'n}ski, I., Gieren, W., Pietrzy{\'n}ski, G., et al.\ 2006, \apj, 648, 375

\bibitem[Soszy{\'n}ski et al.(2007)]{2007AcA....57..201S} Soszy{\'n}ski, I., 
Dziembowski, W.~A., Udalski, A., et al.\ 2007, \actaa, 57, 201 

\bibitem[Soszy{\'n}ski et al.(2008)]{2008AcA....58..163S} Soszy{\'n}ski, I., 
Poleski, R., Udalski, A., et al.\ 2008, \actaa, 58, 163 

\bibitem[Soszy{\'n}ski et al.(2010)]{2010AcA....60...17S} Soszy{\'n}ski, 
I., Poleski, R., Udalski, A., et al.\ 2010, \actaa, 60, 17

\bibitem[Storm et al.(2011)]{2011A&A...534A..95S} Storm, J., Gieren, W., Fouqu{\'e}, P., et al.\ 2011, \aap, 534, A95 

\bibitem[Thompson et al.(2001)]{2001AJ....121.3089T} Thompson, I.~B., Kaluzny, J., Pych, W., et al.\ 2001, \aj, 121, 3089 


\bibitem[Trundle et al.(2007)]{2007A&A...471..625T} Trundle, C., Dufton, P.~L., Hunter, I., et al.\ 2007, \aap, 471, 625 

\bibitem[Udalski et al.(2015)]{2015AcA....65....1U} Udalski, A., Szyma{\'n}ski, M.~K., \& Szyma{\'n}ski, G.\ 2015, \actaa, 65, 1

\bibitem[Ulaczyk et al.(2013)]{2013AcA....63..159U} Ulaczyk, K., 
Szyma{\'n}ski, M.~K., Udalski, A., et al.\ 2013, \actaa, 63, 159 

\bibitem[Urbaneja et al..(2017)]{2017ApJ...} Urbaneja, M., 
Kudritzki, R.-P., , et al.\ 2017, \apj, submitted

\end{thebibliography}
\end{document}